\documentclass[journal,12pt,onecolumn,draft]{IEEEtran}

\usepackage[utf8]{inputenc}
\usepackage[english]{babel}
\usepackage{amsmath}
\usepackage{amssymb}
\usepackage{graphicx}
\usepackage{enumerate}
\usepackage{algorithm}
\usepackage{algpseudocode}

\DeclareMathOperator{\supp}{supp}
\DeclareMathOperator{\wt}{wt}

\newcommand{\F}{\ensuremath{\mathbb{F}}}
\newcommand{\Z}{\ensuremath{\mathbb{Z}}}
\newcommand{\A}{\ensuremath{\mathcal{A}}}
\newcommand{\B}{\ensuremath{\mathcal{B}}}

\makeatletter
\newcommand{\removelatexerror}{\let\@latex@error\@gobble}
\makeatother

\renewcommand{\vec}[1]{#1}

\newtheorem{definition}{Definition}
\newtheorem{proposition}{Proposition}
\newtheorem{theorem}{Theorem}
\newtheorem{lemma}{Lemma}
\newtheorem{corollary}{Corollary}
\newtheorem{remark}{Remark}

\begin{document}
%
\title{On Decoding of Generalized Concatenated Codes and Matrix-Product Codes}
%
%
%

\author{Ferdinand~Blomqvist,
        Oliver~W.~Gnilke,
        and~Marcus~Greferath
\thanks{Ferdinand Blomqvist is with the Department of Mathematics and Systems Analysis, Aalto University.}
\thanks{Oliver W. Gnilke is with the Department of Mathematical Sciences, Aalborg University.}
\thanks{Marcus Greferath is with University College Dublin.}}

\maketitle

\begin{abstract}
  Generalized concatenated codes were introduced in the 1970s by Zinoviev.
  There are many types of codes in the literature that are known by other names that can be viewed as generalized concatenated codes.
  Examples include matrix-product codes, multilevel codes and generalized cascade codes.
  Decoding algorithms for generalized concatenated codes were developed during the 1970s and 1980s.
  However, their use does not appear to be as widespread as it should, especially for codes that are known by other names but can be viewed as generalized concatenated codes.

  In this paper we review the decoding algorithms for concatenated codes, generalized concatenated codes and matrix-product codes, and clarify the connection between matrix-product codes and generalized concatenated codes.
  We present a small improvement to the decoding algorithm for concatenated codes.
  We also extend the decoding algorithms from errors-only decoders to error-and-erasure decoders.
  Furthermore, we improve the upper bound on the computational complexity of the decoding algorithm in the case of matrix-product codes where the generator matrix for the inner code is non-singular by columns.
\end{abstract}


%
\IEEEpeerreviewmaketitle

\section{Introduction}
\label{sec:intro}
Matrix-product codes form a class of generalized concatenated codes (GCC) and were introduced by Blackmore and Norton \cite{blackmore2001matrix}.
They can also be seen as generalizations of the Plotkin sum construction, which is also known as the $(u\ |\ u+v )$ construction.
Decoding algorithms for matrix-product codes were considered by Hernando \emph{et. al.} in \cite{hernando2009construction} and \cite{hernando2013decoding}.
These decoding algorithms can correct all error patterns up to half the minimum distance of the code, but places restrictions on the choice of the component codes.
In addition, the algorithms are computationally intensive and therefore intractable in many cases.

By utilizing Forney's \emph{generalized minimum distance} (GDM) decoding we devised an efficient decoding algorithm for matrix-product codes that can correct all error patterns of weight less than half the minimum distance, and many patterns of larger error weight.
In the final stages of our work we found out that most of our work had already been done in the context of GCC already in the 70s and 80s.
Most of the old research on GCC is only published in Russian, and the community working on matrix-product codes seems to be mostly unaware of its existence.
Therefore this article serves the purpose of raising awareness of the existing research.
In addition, this is, to the best of our knowledge, the first comprehensive treatment of this subject available in English.
The treatments of the subject available in English include \cite{ericson1986simple,bossert1988decoding}, while \cite{zyablov1999introduction} serves as a good introduction to concatenated codes and GCC, but it only briefly mentions the basics of the decoding algorithm.
The decoding of GCC up to half the minimum distance is due to Blokh and Zyablov \cite{blokh1974coding}.
Improvements are due to Zinoviev \cite{zinov1981generalized} and Bossert \cite{bossert1988decoding}.

The paper is organized as follows.
Section \ref{sec:prelim} establishes the notation and presents the necessary preliminaries.
Concatenated codes, GCC and matrix-product codes are presented in Sections \ref{sec:cc},\ref{sec:gcc}, and \ref{sec:mpcodes} respectively.
Sections \ref{sec:decodecc}, \ref{sec:decodegcc} and \ref{sec:decodemp} deal with decoding of said codes.
In Section \ref{sec:examples} we break down the decoding algorithm for two simple matrix-product codes.
Finally, error-and-erasure decoding is discussed in Section \ref{sec:erasmp}, and in Section \ref{sec:beyondmd} we discuss different methods for increasing the error correction capabilities of the decoding algorithm.

\section{Preliminaries and notation}
\label{sec:prelim}
For $n \in \Z,\ n \geq 1$, define $[n] := \{i \in \Z\ |\ 1 \leq i \leq n \}$.
For a prime power $q$, $\F_{q}$ denotes the finite field with $q$ elements.
The support of $\vec{c} \in \F_{q}^n$ is defined as
\begin{equation*}
  \supp(\vec{c}) := \{ i \in [n]\ |\ c_i \neq 0 \},
\end{equation*}
and $\wt(\vec{c}) = |\supp(\vec{c})|$ denotes the Hamming weight of $\vec{c}$.
Given $E \subset [n]$, $$\wt_{E}(\vec{c}) := |\supp(\vec{c}) \setminus E |$$ denotes the Hamming weight of $\vec{c}$ punctured at the coordinates in $E$.
The set of all $M \times N$ matrices with entries in $\F_q$ is denoted by $\F_q^{M \times N}$.
The power set of a set $S$ is denoted by $\mathcal{P}(S)$.
$x_i$ denotes the $i$-th coordinate of $\vec{x}$.

A linear code over $\F_q$ with length $n$, dimension $k$ and minimum distance $d$ is said to be an $[n, k, d]_q$ code.
The field size will be omitted when it is clear from the context.

\subsection{Generalized minimum distance decoding}
\label{sec:gmd}
For the reminder of the section, let $C \subset \F_{q}^{n}$ be a code with minimum distance $d$, $\vec{c} \in C$ and $\vec{r} \in \F_{q}^{n}$.
The following results are well known.
\begin{theorem}
  \label{thm:unique1}
  Given $C$ and $\vec{r}$, there is at most one codeword $\vec{c} \in C$ such that
  \begin{equation}
    \label{eq:mddcond}
    2 \wt(\vec{r} - \vec{c}) < d.
  \end{equation}
\end{theorem}
\begin{theorem}
  \label{thm:unique2}
  Given $C$, $\vec{r}$ and $E \subset [n]$, there is at most one codeword $\vec{c} \in C$ such that
  \begin{equation}
    \label{eq:elias}
    2 \wt_{E}(\vec{r} - \vec{c}) + |E| < d.
  \end{equation}
\end{theorem}

Theorem \ref{thm:unique1} is the familiar statement that the spheres of Hamming radius $(d - 1) / 2$ centred at the codewords do not intersect.
Theorem \ref{thm:unique2} is the analogous statement for error-and-erasure decoding.
A errors-only decoder that only corrects words that satisfy \eqref{eq:mddcond} is called a \emph{minimum distance} decoder.

In errors-only decoding the decoder is only given the received word.
In order to ease the task of the decoder, one can also give the decoder information about the reliability of each symbol.
If we give the decoder information which symbols are erased (unreliable) and not erased (reliable), then we are performing error-and-erasure decoding.
It follows from Theorem \ref{thm:unique2} that an error-and-erasure decoder that decodes every received word that satisfies \eqref{eq:elias}, and rejects all other received words, can be described by a map
\begin{equation*}
  \sigma: \F_q^n \times \mathcal{P}([n]) \to C \cup \{ \otimes \}
\end{equation*}
where $(\vec{r}, E)$ is mapped to the (unique) closest codeword, or to $\otimes$ if \eqref{eq:elias} is not satisfied.
Hence $\sigma(\vec{r}, E) = \vec{c}$ if and only if $2 \wt_{E}(\vec{r} - \vec{c}) + |E| < d$.

There is of course no reason to restrict the supplied symbol reliability information to only two reliability classes.
In $1965$ Forney considered what happens if the received symbols are classified into $J$ reliability classes and arrived at what he named GMD decoding \cite{forney1965concatenated}.
For shorter notation, define $I_E = \supp(\vec{r} - \vec{c})$ and $I_C = [n] \setminus I_E$.
Furthermore, let $\vec{\alpha} \in [0,1]^{n}$ be the supplied reliability weight vector, meaning that $\alpha_i$ is the reliability of $r_i$.
A smaller reliability weight means that the symbol is considered less reliable.
We have the following theorem.
\begin{theorem}[Forney \cite{forney1965concatenated}]
  \label{thm:forney1}
  Given $C, \vec{r}$ and $\vec{\alpha}$ there is at most one codeword $\vec{c} \in C$ such that
  \begin{equation}
    \label{eq:gmdcond}
    \sum_{i \in I_C} (1 - \alpha_i)
    + \sum_{i \in I_E} (1 + \alpha_i) < d.
  \end{equation}
\end{theorem}

Suppose we have $J$ reliability classes with corresponding reliability weights $a_j$, $1 \leq j \leq J$, and $a_j \leq a_k$ for $j < k$.
Each symbol is put into one of these $J$ reliability classes.
Let $E_j(\alpha) := \{ i \in [n]\ |\ \alpha_i \leq a_j \}$, $1 \leq j \leq J$, and $E_0(\alpha) := \emptyset$.
We will omit the parameter $\alpha$ whenever it is clear from the context.

\begin{theorem}[Forney \cite{forney1965concatenated}]
  \label{thm:forney2}
  If \eqref{eq:gmdcond} holds, then there exists $0 \leq j < J$, such that
  \begin{equation}
    \label{eq:erasurecondj}
    2 \wt_{E_j}(\vec{r} - \vec{c}) + |E_j| < d.
  \end{equation}
\end{theorem}
\begin{remark}
  Forney showed that if \eqref{eq:gmdcond} holds, then there exists $0 \leq j \leq J$, such that \eqref{eq:erasurecondj} is satisfied.
  However, \eqref{eq:erasurecondj} cannot hold for $j = J$, since $|E_J| = n > d$.
\end{remark}
We say that a $j$ is \emph{correct} if \eqref{eq:erasurecondj} holds for this $j$.
Theorem \ref{thm:forney2} shows that GMD decoding can be implemented with an error-and-erasure decoder and Theorem \ref{thm:forney1} can be used to check if the chosen $j$ was correct.

It is clear that $\sigma(\vec{r},E_j) \neq \vec{c}$ if $|E_j| \geq d$, and hence we immediately get an upper bound on the maximum number of different $j$ we need to try, namely at most $d$ different ones.
It is, however, possible to strengthen this bound.
\begin{theorem}
  \label{thm:evenearsure}
  Let $F_1 \subset F_2 \subset [n]$ be such that $|F_2| = |F_1| + 1$, and suppose $d - |F_1|$ is even.
  If $\sigma(\vec{r}, F_1) \neq \otimes$, then
  \begin{equation*}
    \sigma(\vec{r}, F_1) = \sigma(\vec{r}, F_2).
  \end{equation*}
\end{theorem}
\begin{IEEEproof}
  We prove this by contradiction.
  Let $\vec{x_j} = \sigma(\vec{r}, F_j), j \in \{1,2\}$, and suppose that $\vec{x_1} \neq \vec{x_2}$.
  We have $2 \wt_{F_1}(\vec{r}-\vec{x_1})+ |F_1| < d$ by assumption, and thus $2 \wt_{F_2}(\vec{r}-\vec{x_1}) + |F_2| \geq d$ by Theorem \ref{thm:unique2}.
  Furthermore, $\wt_{F_2}(\vec{r} - \vec{c}) \leq \wt_{F_1}(\vec{r} - \vec{c})$, and therefore we must have $\wt_{F_2}(\vec{r}-\vec{x_1}) = \wt_{F_1}(\vec{r}-\vec{x_1})$.
  It follows that $2 \wt_{F_1}(\vec{r}-\vec{x_1}) + |F_1| = d - 1$, which contradicts the assumption that $d - |F_1|$ is even.
\end{IEEEproof}

We call the act of running the decoder for $C$ with one erasure set and then checking if Theorem \ref{thm:forney1} holds for the decoded word a \emph{trial}.
\begin{corollary}
  \label{cor:trials}
  At most $\lfloor (d+1) / 2 \rfloor$ trials are required to decode any received word $r$ that satisfies Theorem \ref{thm:forney1}.
\end{corollary}

This result was also noted by Forney, but he presented no formal proof.
We will end this section with a few helpful observations.
These observations do not, unfortunately, affect the worst case complexity of the GMD decoder.
They do, however, provide a way to eliminate unnecessary trials after we know the received word.

If there are no symbols with reliability weight $a_j$, then $E_{j} = E_{j-1}$, and hence we do not need to run the trial for this value of $j$.
Furthermore, by Theorem \ref{thm:evenearsure}, if $j$ is such that $d - |E_j|$ is even and $|E_{j+1}| = |E_j| + 1$, then this trial can be omitted.
We call $j>0$ \emph{viable} if it does not satisfy either of the two previous conditions.
The GMD decoder is described formally as Algorithm \ref{alg:gmd}.
\begin{figure}[htbp]
  \removelatexerror
  \begin{algorithm}[H]
    \caption{GMD decoding for the component codes}
    \label{alg:gmd}
    \begin{algorithmic}[1]
      \Procedure{gmd-decode}{$C,r,\alpha$}
	\For{$j=0$ to $J-1$}
	  \If{$j$ is viable or $j=0$}
	    \State $c := \sigma_{C}(r, E_j(\alpha))$
	    \If{$c \neq \otimes$ and \eqref{eq:gmdcond} is satisfied}
	      \State \Return $c$
	    \EndIf
	  \EndIf
	\EndFor
	\State \Return Failure
      \EndProcedure
    \end{algorithmic}
  \end{algorithm}
\end{figure}

\subsection{Non-singular by columns matrices}
\begin{definition}
  For $\vec{A} \in \F_q^{M \times N}$ and $t \in [M]$, let $\vec{A}_t$ denote the $t$ first rows of $\vec{A}$, and for $1 \leq j_1 \leq \dots \leq j_i \leq N$, $\vec{A}(j_1, \dots, j_i)$ denotes the matrix consisting of columns $j_1, \dots, j_i$ of $\vec{A}$.
  We call $\vec{A}$ \emph{non-singular by columns} (NSC) if $\vec{A}_t(j_1, \dots, j_t)$ is non-singular for all $t \in [M]$ and all $1 \leq j_1 \leq \dots \leq j_t \leq N$.
\end{definition}

We call a matrix \emph{triangular} if it is a column permutation of an upper-triangular matrix.

\begin{proposition}
  \label{prop:mds}
  If $\vec{A}$ is NSC, then, for all $1 \leq t \leq M$, the linear code generated by $\vec{A}_t$ is MDS.
\end{proposition}
\begin{IEEEproof}
  Recall that a linear code generated by an $t \times N$ matrix is MDS if and only if any $t$ columns of the generator matrix are linearly independent.
  Considering $\vec{A}_t$, we find that,
  since $\vec{A}$ is NSC, the matrix $\vec{A}_t(j_1, \dots, j_t)$ is non-singular for all $1 \leq j_1 < \dots < j_t \leq N$.
  This implies that any $t$ columns of $\vec{A}_t$ are linearly independent, and thus the code generated by $\vec{A}_t$ is MDS.
\end{IEEEproof}

\section{Concatenated codes}
\label{sec:cc}
Let $\A$ be a code over $\F_{q^s}$ of length $M$ and minimum distance $d_{a}$.
Let $X$ be the message set for $\A$, and let $\tau: X \to \F_{q^s}^M$ be an encoding function for $\A$.
The code $\A$ is called the \emph{outer code}.
Fix a basis for $\F_{q^s}$ over $\F_q$.
Then elements of $\F_{q^s}$ can be represented as vectors of length $s$ over $\F_{q}$.
Thus elements of $\A$ can be represented as $M \times s$ matrices over $\F_{q}$, and we will use this representation throughout this section.

Let $\B_i$, $i \in [M]$, be linear $[N, K, d_{b}]$ codes over $\F_{q}$, where $K = ks$ for some integer $k$.
Furthermore, let $\sigma_i: \F_{q}^{K} \to \F_{q}^{N}, i \in [M]$,
be the encoding functions associated with $\B_i$.
The $\B_i$ are called the \emph{inner codes}.

Let $\vec{Y} = (x_1, \dots, x_k),\ x_i \in X$, be the message we want to encode.
The encoded message is

\begin{equation*}
  \vec{W} =
  \begin{bmatrix}
    \sigma_1(\vec{V}_1) \\
    \vdots \\
    \sigma_{M}(\vec{V}_{M})
  \end{bmatrix},
\end{equation*}
where
\begin{equation*}
  \vec{V} =
  \begin{bmatrix}
    \tau(x_1) & \dots & \tau(x_{k})
  \end{bmatrix}.
\end{equation*}
Or in other words, to encode $Y$ with the concatenated code we first encode each of the $x_i$ with the outer code, and use the result to build the $M \times K$ matrix $V$.
Then we encode each row of $V$ with its corresponding inner code.
More precisely, the $i$-th row is encoded with $\B_i$.

We denote this concatenated code by $\A \{ \B_i \}_{i \in [k]}$, or simply $\A \B$ in the case of one inner code.
It is clear that $\vec{W} \in \F_{q}^{M \times N}$, and hence the concatenated code has length $MN$.
The size of the concatenated code is $|\A|^k$.
The minimum Hamming distance $d$ of the concatenated code satisfies $d \geq d_{a} d_{b}$.
To see this, suppose that $\vec{Y} \neq \vec{Y}'$.
Then $\vec{V}$ and $\vec{V}'$ differ in at least $d_{a}$ rows, which implies that $\wt(\vec{W} - \vec{W}') \geq d_{a} d_{b}$.

If $\A$ is linear with dimension $k_a$, then the resulting concatenated code is also linear over $\F_q$, and has dimension $k_a \cdot K$, since, in this case, $|\A|^k = ((q^s)^{k_a})^k = q^{k_a K}$.

Since, $\B_i$ is linear, $\sigma_i$ can be written in terms of the generator matrix of $\B_i$.
Let $\vec{B_i}$ be a generator matrix of $\B_i$.
Then $\sigma_i(\vec{V}_i) = \vec{V}_i \vec{B_i}$.
If all $\B_i$ are the same, then we let $\vec{B} := \vec{B_i}$ and we can simply write $\vec{W} = \vec{V} \vec{B} $.
For most practical purposes it seems reasonable to let all the inner codes be the same.

One example of concatenated coding is so-called product codes (also known as iterative codes) introduced by Elias in 1954 \cite{elias1954error}.
We refer to \cite{zyablov1999introduction} for more examples.

\section{Generalized concatenated codes}
\label{sec:gcc}
We will now consider GCC.
In order to ease the presentation we will restrict to the case where we only have one inner code.
It is straight forward to extend to the case of several inner codes, but the notation quickly gets tedious.
In addition, the use of several inner codes adds complexity for no apparent gain, since the main parameters of the generalized concatenated code does not change when using several inner codes.

Let $\A_i$, $i \in [k]$, be codes over $\F_{q^{s_i}}$ of length $M$ and minimum distance $d_{a,i}$.
Let $X_i$ be the message set for $\A_i$, and let $\tau_i: X_i \to \F_{q^{s_i}}^M$ be an encoding function for $\A_i$. The $\A_i$ are called the \emph{outer codes}. Elements of $\A_i$ will again be represented as $M \times s_i$ matrices over $\F_{q}$.

Let $\B$ be a linear $[N, K, d_{b}]$ code over $\F_{q}$, where $K = \sum_{j=1}^k s_j$.
Furthermore, let $\sigma: \F_{q}^{K} \to \F_{q}^{N}, i \in [M]$,
be the encoding function associated with $\B$.
The code $\B$ is called the \emph{inner code}.

Let $\vec{Y} = (x_1, \dots, x_k), x_i \in X_i$, be the message we wish to encode.
The encoded message is
\begin{equation*}
  \vec{W} =
  \begin{bmatrix}
    \sigma(\vec{V}_1) \\
    \vdots \\
    \sigma(\vec{V}_{M})
  \end{bmatrix},
\end{equation*}
where
\begin{equation*}
  \vec{V} =
  \begin{bmatrix}
    \tau_{1}(x_1) & \cdots & \tau_{k}(x_{k})
  \end{bmatrix}.
\end{equation*}
This code is called a $k$-th order GCC.
The length of the GCC is again $MN$, and the size is $|\A_1| \cdots |\A_k|$.
The true minimum distance $d$ of the code is not known in most cases, but it can be lower bounded.
We denote this GCC by $[\A_1, \dots, \A_k] \B$.

If the outer codes are linear with dimension $k_{a,i}$, then the generalized concatenated code is also linear over $\F_q$, and has dimension $\sum_{i \in [k]} k_{a,i} \cdot s_i$
To see this, simply note that
\begin{equation*}
  \prod_{i \in [k]} |\A_i| = \prod_{i \in [k]} ((q^{s_i})^{k_{a,i}} = q^{\sum_{i \in [k]} s_i \cdot k_{a,i}}.
\end{equation*}

We will now derive the lower bound for the minimum distance.
First, however, we need to introduce some notation.
Since, $\B$ is linear, $\sigma$ can be written in terms of the generator matrix of $\B$.
For this purpose, let $\vec{B}$ be a generator matrix of $\B$.
Then $\vec{W} = \vec{V} \vec{B}$.
For $i \in [k]$, let $\B^{(i)}$ denote the subcode of $\B$ generated by the first $b_i := \sum_{j=1}^i s_j$ rows of $\vec{B}$.
This gives us a collection of nested codes such that
\begin{equation*}
  \B^{(1)} \subset \B^{(2)} \subset \dots \subset \B^{(k)} = \B.
\end{equation*}
The minimum distance of $\B^{(i)}$ is denoted by $d_{b,i}$.

Suppose $\vec{Y} \neq \vec{Y}'$.
Then $\tau_i(x_i) \neq \tau_i(x'_i)$ for at least one $i$, and let $j$ be the largest such $i$.
The minimum distance of $\A_j$ is $d_{a,j}$ and hence $\vec{V}$ and $\vec{V}'$ differ in at least $d_{a,j}$ rows.
Therefore $\vec{V}-\vec{V}'$ is non-zero in at least $d_{a,j}$ rows, and in addition, the last $N - b_j$ columns of $\vec{V}-\vec{V}'$ are zero.
Thus, every row of $\vec{W}-\vec{W}' = (\vec{V}-\vec{V}') \vec{B}$ is a codeword of $\B^{(j)}$, and it follows that $\wt(\vec{W}-\vec{W}') \geq d_{a,j} \cdot d_{b,j}$.
This holds for every $j \in [k]$, and hence
\begin{equation*}
  d \geq \wt(\vec{W}-\vec{W}') \geq \min_{i \in [k]} \left[ d_{a,i} \cdot d_{b,i} \right].
\end{equation*}
We see that the minimum distance properties of GCC depend on the chosen encoding function for the inner code.
Thus the task of designing a good generalized concatenated code requires more than finding good outer and inner codes.
We also need to find a generator matrix for the inner code that generates a system of nested codes that have good minimum distance properties.
This is, according to \cite{zyablov1999introduction}, one of the main problems for generalized concatenated systems.

\section{Matrix-product codes}
\label{sec:mpcodes}
Matrix-product codes were introduced by Blackmore and Norton \cite{blackmore2001matrix}.
They form a subclass of the class of GCC.
\begin{definition}
  Let $\vec{B} \in \F_{q}^{k \times N}$, and $\A_1, \dots, \A_k$ be codes over $\F_{q}$ of length $M$.
  The \emph{matrix-product code} $[\A_1 \dots \A_k] \cdot \vec{B}$ is the set of all matrix products $[\vec{a_1} \dots \vec{a_k}] \cdot \vec{B}$, where $\vec{a_i} \in \A_i, i \in [k]$ is a column vector.
\end{definition}
We see that matrix-product codes are GCC where $s_i = 1$ for all $i \in [k]$.
The codewords in $C = [\A_1 \dots \A_k] \cdot \vec{B}$ are of the form
\begin{equation*}
  [\vec{a_1} \dots \vec{a_k}] \cdot \vec{B}
  =
  \begin{bmatrix}
    \sum_{i=1}^{k} b_{i1} \vec{a_i} & \dots & \sum_{i=1}^{k} b_{i N} \vec{a_i}
  \end{bmatrix}.
\end{equation*}
It is clear that $C$ has length $M N$, and the following Theorem was proven in \cite{blackmore2001matrix}.
The length, size, and minimum distance bound of the code follows directly from the corresponding properties of GCC.
However, if $\vec{B}$ is NSC and triangular, then the exact minimum distance of the code is known.
\begin{theorem}[Blackmore, Norton \cite{blackmore2001matrix}]
  If $\vec{B}$ is NSC and $C = [\A_1 \dots \A_k] \cdot \vec{B}$, then
  \begin{enumerate}[(i)]
    \item $|C| = |\A_1| \cdots |\A_k|$;
    \item $d \geq d^{*} = \min_{i \in [k]} \left[ d_{a,i} \cdot (N-(i-1)) \right]$;
    \item if $\vec{B}$ is also triangular, then $d = d^*$.
  \end{enumerate}
  \label{thm:minimumdistance}
\end{theorem}
Since matrix-product codes are also GCC, we know that if the codes $\A_i, i \in [k]$, are linear, then $C = [\A_1 \dots \A_k] \cdot \vec{B}$ is also linear.
In addition, the dimension of $C$ is the now simply the sum of the dimensions of the $\A_i$.

Well known examples of matrix-product codes include the $(u\ |\ u + v)$ and
$(u + v + w\ |\ 2u + v\ |\ u)$ constructions.
If we choose
\begin{equation*}
  \vec{B} =
  \begin{bmatrix}
    1 & 1 \\
    0 & 1
  \end{bmatrix}
  \quad \text{or} \quad
  \vec{B} =
  \begin{bmatrix}
    1 & 2 & 1 \\
    1 & 1 & 0 \\
    1 & 0 & 0
  \end{bmatrix}
  ,
\end{equation*}
then we obtain the $(u\ |\ u + v)$ or $(u + v + w\ |\ 2u + v\ |\ u)$ construction, respectively.

\section{Decoding of concatenated codes}
\label{sec:decodecc}
We will now show how to decode concatenated codes.
For simplicity we will only consider the case where all the inner codes are the same.
It is simple to adapt the decoding algorithm to the case of several inner codes.

It is non-trivial to devise a decoding algorithm for concatenated codes that can correct all error patterns of weight less than half the designed minimum distance.
The algorithm itself is quite simple though.
Consider the concatenated code $\A \B$, and let $\vec{R} = \vec{W} + \vec{E}$ be the received word.
The first step is to decode every row of $\vec{R}$ with a minimum distance decoder for the inner code $\B$.
Denote the result of this operation with $\widehat{\vec{W}}$, and let $\widehat{\vec{E}} = \vec{R} - \widehat{\vec{W}}$.
For any matrix $\vec{A}$, let $\vec{A}_i$ and $\vec{A}^i$ denote the $i$-th row and column of $\vec{A}$, respectively.
We will consider $\vec{V}$ as a matrix with elements in $\F_{q^{s}}$, and thus $\vec{V}$ has $k$ columns.

Define
\begin{equation*}
  w_j :=
  \begin{cases}
    2 \wt(\widehat{\vec{E}}_j), & \text{if }  2 \wt(\widehat{\vec{E}}_j) < d_b, \\
    d_b, & \text{in case of decoding failure},
  \end{cases}
\end{equation*}
and, for all $ \in [M]$, assign the $j$-th row the reliability weight $\alpha_j := (d_b - w_j) / d_b$.

For every row with a non-zero reliability weight, \emph{i.e.}, the rows where the row decoder succeeded, find the message $\widehat{\vec{V}}_j$ that corresponds to $\widehat{\vec{W}}_j$.
Thus $\sigma(\widehat{\vec{V}}_j) = \widehat{\vec{W}}_j$.
For rows with reliability weight $\alpha_j = 0$, set $\widehat{\vec{V}}_j = 0$ (or some other message).
Now, for $i \in [k]$, decode $\widehat{\vec{V}}^i$ with a GMD decoder for the outer code $\A$ using $\alpha$ as the reliability weight vector.
Let $\Phi$ denote this decoding algorithms.

\begin{theorem}
  The decoder $\Phi$ can correct all error patterns $\vec{E}$ that satisfy
  \begin{equation}
    \label{eq:errorcond}
    \sum_{i \in [M]} \min\left\{ \wt(\vec{E}_i),\ d_b \right\} < \frac{d_a d_b}{2}.
  \end{equation}
  \label{thm:errorscorrect}
\end{theorem}
\begin{IEEEproof}
  We only need to prove that the generalized minimum distance criterion \eqref{eq:gmdcond} holds for $\widehat{\vec{V}}^i,\ i \in[k]$, after the row decoding.
  Let $I_C \subset [M]$ denote all rows that were correctly decoded, and define $I_E := [M] \setminus I_C$.
  Suppose $j \in I_C$.
  Then
  \begin{IEEEeqnarray*}{rCl}
    2 \min \left\{ \wt(\vec{E}_j),\ d_b \right\}
    &=& 2 \wt(\vec{E}_j)
    = 2 \wt(\widehat{\vec{E}}_j)
    = d_b - \left( d_b - w_j \right)
    = (1 - \alpha_j) d_b.
  \end{IEEEeqnarray*}
  If $j \in I_E$, and we do not have a decoding failure for that row, then
  \begin{IEEEeqnarray*}{rCl}
    2 \min \left\{ \wt(\vec{E}_j),\ d_b \right\}
    & \geq & 2 d_b - 2 \wt(\widehat{\vec{E}}_j)
    = d_b + \left( d_b - w_j \right)
    = (1 + \alpha_j) d_b.
  \end{IEEEeqnarray*}
  On the other hand, if we have a decoding failure for the $j$-th row, then $\alpha_j = 0$, and hence
  \begin{IEEEeqnarray*}{rCl}
    2 \min \left\{ \wt(\vec{E}_j),\ d_b \right\}
    & \geq & d_b = (1 + \alpha_j) d_b.
  \end{IEEEeqnarray*}

  Now, assume that \eqref{eq:errorcond} holds.
  Then,
  \begin{IEEEeqnarray*}{rCl}
    \sum_{i \in I_C} (1 - \alpha_i) d_b
    + \sum_{i \in I_E} (1 + \alpha_i) d_b
    & \leq &
    \sum_{j \in [M]} 2 \min \left\{ \wt(\vec{E}_j),\ d_b \right\}
    < d_a d_b,
  \end{IEEEeqnarray*}
  which shows that \eqref{eq:gmdcond} holds for all columns of $\widehat{\vec{V}}$.
\end{IEEEproof}
\begin{corollary}
  \label{cor:errorscorrect}
  The decoder $\Phi$ can correct all error patterns $\vec{E}$ that satisfy $2 \wt(\vec{E}) < d_a d_b$.
\end{corollary}

We have already seen that the decoder can correct the maximum number of random errors.
It can also correct many bursty error patterns with much higher weight.
We call a row \emph{bursty} if it has at least $d_b / 2$ errors.
\begin{corollary}
  \label{cor:bursty}
  Let $b$ denote the number of bursty rows, and $t$ the remaining number of errors.
  The decoder $\Phi$ can correct any error pattern such that $2 (t + b d_b) < d_a d_b$.
\end{corollary}
\begin{IEEEproof}
  If the $i$-th row is not bursty, then $\min \left\{ \wt(\vec{E}_i),\ d_b \right\} = \wt(\vec{E}_i)$.
  On the other hand, if the $i$-th row is bursty, then $\min \left\{ \wt(\vec{E}_i),\ d_b \right\} \leq d_b$, and hence the result follows.
\end{IEEEproof}

We have $J := \lfloor (d_b + 1) / 2 \rfloor + 1$ reliability classes.
Thus, by Corollary \ref{cor:trials}, $\widehat{\vec{V}}^i$ can be recovered in at most $\min \{J, \lfloor (d_a + 1) / 2 \rfloor \}$ trials.
However, due to the design of the reliability weights, we can lower this bound slightly.
More, precisely, the GMD decoder can always consider any row with zero reliability weight as an erasure.

For $j \in [J]$, let $F_j := \{ i \in [M]\ |\ \alpha_i \leq a_j \}$, where the $a_j$ are the weights corresponding to each reliability class ordered from the smallest to the largest.
Furthermore, let $F_0 = \emptyset$.
From the proof of Theorem \ref{thm:errorscorrect} and Theorem \ref{thm:forney2} we know that there exists $0 \leq j < J$ such that
\begin{equation}
  \label{eq:gmdrows}
  2 | I_E \setminus F_j | + |F_j| < d_b.
\end{equation}
Now, suppose that \eqref{eq:gmdrows} holds for $j = 0$.
Then, since $F_1 \subset I_E$, it follows that
\begin{equation*}
  2 | I_E \setminus F_1 | + |F_1| \leq 2 |I_E| < d_b.
\end{equation*}
Therefore, we can skip the trial where $j = 0$.
This means that we only need at most
\begin{equation*}
  m := \left\lfloor \frac{\min \{d_a, d_b \} + 1}{2} \right\rfloor
\end{equation*}
trials to decode any $\widehat{\vec{V}}^i$.

\subsection{Improving the algorithm}
The obvious way to improve the algorithm is to limit the number of trials that the GMD decoder has to run.
As can be seen in the proof of Theorem \ref{thm:errorscorrect}, there is one $ 1 \leq j < J$ such that the GMD decoder will decode any $\widehat{\vec{V}}^i$ correctly with the erasure set $F_j$.
This can be leveraged to reduce the number of times the decoder for the outer code $\A$ has to be run.

Instead of having the GMD decoder start from $j = 1$ for every $\widehat{\vec{V}}^i$, we can do the following:
Start from $j = 1$ when decoding $\widehat{\vec{V}}^1$, but when decoding $\widehat{\vec{V}}^{i+1}$ start from the $j$ that was used to successfully decode $\widehat{\vec{V}}^i$.
This way the upper bound for the total number of times we decode with the outer code is reduced from $km$ to $k + m - 1$.

We presented this improvement in the context of product codes \cite{blomqvist2020}, but, to the best of our knowledge, this improvement has not been considered for concatenated codes.
The only downside of this modification is that the inherent parallelism of the algorithm is lost.

\section{Decoding of generalized concatenated codes}
\label{sec:decodegcc}
A $k$-th order GCC can be decoded by applying the decoding algorithm for concatenated codes $k$-times.

Consider the GCC $[\A_1, \dots, \A_k] \B$, and let $\vec{R} = \vec{W} + \vec{E}$ be the received word.
Furthermore, let $\vec{B}$ be a generator matrix for $\B$.
We can use the decoding algorithm for the concatenated code $\A_k \B^{(k)}$ to recover $\vec{V}^k$.
Then we can cancel out the contribution of $\vec{V}^k$ to $\vec{R}$, by letting
\begin{equation*}
  \vec{R}' := \vec{R} -
  \begin{bmatrix}
    \vec{0} & \dots & \vec{0} & \vec{V}^k
  \end{bmatrix}
  \vec{B}
  =
  \begin{bmatrix}
    \vec{V}^1 & \dots & \vec{V}^{k-1} & \vec{0}
  \end{bmatrix}
  \vec{B}
  + \vec{E}
\end{equation*}
Hence, $\vec{R}' = \vec{W}' + \vec{E}$, where $\vec{W}'$ is a codeword in $[\A_1, \dots, \A_{k-1}] \B^{(k-1)}$.
Thus, $\vec{R}$ can be decoded with decoders for the concatenated codes $\A_i \B^{(i)},\ i \in [k]$.
The decoding algorithm is described more formally as Algorithm \ref{alg:gccbasic}, and $\vec{B}^{-1}$ denotes a right inverse of $\vec{B}$ in the algorithm description.
\begin{figure}[htbp]
  \removelatexerror
  \begin{algorithm}[H]
    \caption{Decoding of generalized concatenated codes}
    \label{alg:gccbasic}
    \begin{algorithmic}[1]
      \Procedure{gccdecode}{$[\A_{1}, \dots, \A_{k}] \B,\vec{R}$}
	\State $\widehat{\vec{W}}, \alpha := $ \Call{rowdecode}{$\B, \vec{R}$}
	\State $\widehat{\vec{V}} := \widehat{\vec{W}} \vec{B}^{-1}$
	\State $\vec{V}^k := $ \Call{gmd-decode}{$\A_k, \widehat{\vec{V}}^k, \alpha$}
	\State $\vec{R}' := \vec{R} - \begin{bmatrix} \vec{0} & \dots & \vec{0} & \vec{V}^k \end{bmatrix} \vec{B}$
	\If{$k > 1$}
	  \State $\vec{W}' := $ \Call{gccdecode}{$[\A_{1}, \dots, \A_{k-1}] \B^{(k-1)}, \vec{R}'$}
	  \State \Return $\vec{W}' + \begin{bmatrix} \vec{0} & \dots & \vec{0} & \vec{V}^k \end{bmatrix} \vec{B}$
	\Else
	  \State \Return $\begin{bmatrix} \vec{0} & \dots & \vec{0} & \vec{V}^k \end{bmatrix} \vec{B}$
	\EndIf
      \EndProcedure
      \State
      \Procedure{rowdecode}{$C, \vec{R}$}
	\State Decode every row of $\vec{R}$ with the code $C$.
      \EndProcedure
    \end{algorithmic}
  \end{algorithm}
\end{figure}
Let $\Psi$ denote the decoding algorithm outlined as Algorithm \ref{alg:gccbasic}.
We have the following results which are analogous to Theorem \ref{thm:errorscorrect} and Corollary \ref{cor:errorscorrect}.
\begin{theorem}
  The decoder $\Psi$  can correct all error patterns $\vec{E}$ that satisfy
  \begin{equation}
    \label{eq:gccerrorcond}
    \sum_{j \in [M]} \min\left\{ \wt(\vec{E}_j),\ d_{b,1} \right\} < \frac{d^*}{2},
  \end{equation}
  where $d^*$ is the designed minimum distance of the code.
  \label{thm:gccerrorscorrect}
\end{theorem}
\begin{IEEEproof}
  Recall that $d^* = \min_{i \in [k]} d_{a,i} d_{b,i}$, and hence
  \begin{equation*}
    \sum_{j \in [M]} \min\left\{ \wt(\vec{E}_j),\ d_{b,i} \right\}
    \leq \sum_{i \in [M]} \min\left\{ \wt(\vec{E}_i),\ d_{b,1} \right\} < \frac{d^*}{2} \leq \frac{d_{a,i} \cdot d_{b,i}}{2}
  \end{equation*}
  for all $i \in [k]$.
  Thus, $V^{i}$ will be correctly decoded with the decoder for $\A_{i} \B^{(i)}$, for every $i \in [k]$.
\end{IEEEproof}
\begin{corollary}
  The decoder $\Psi$ can correct all error patterns $\vec{E}$ that satisfy $2 \wt(\vec{E}) < d^*$, where $d^*$ is the designed minimum distance of the code.
\end{corollary}

In the context of GCC we call a row \emph{bursty} if it has at least $d_{b,1} / 2$ errors.
\begin{corollary}
  \label{cor:gccbursty}
  Let $b$ denote the number of bursty rows, and $t$ the remaining number of errors.
  The decoder $\Psi$ can decode any error pattern such that $2 (t + b d_{b,1}) < d^*$.
\end{corollary}
\begin{IEEEproof}
  The proof is analogous to the proof of Corollary \ref{cor:bursty}.
\end{IEEEproof}

The complexity of this algorithm is easy to describe.
The decoder for the inner code $\B^{(i)}$ is run $M$ times, and the decoder for the outer code $\A_i$ is run at most $\lfloor (\min\{d_{a,i}, d_{b,i} \} + 1) / 2 \rfloor$ times.

The algorithm can be improved by leveraging the fact that the $\B^{(i)}$ are nested, and that the error pattern $\vec{E}$ stays the same during all rounds of decoding.
We call the process of recovering $\vec{V}^i$ with the decoder for $\A_i \B^{(i)}$ the $i$-th \emph{round} of decoding.
We have chosen this slightly non-intuitive convention since it makes the notation easier.
During the $i$-th round of decoding we decode all the rows with $\B^{i}$, and during the next round we decode the rows again with $\B^{(i-1)}$.
However, the error vector for each row remains the same and combining this with the fact that $\B^{(i-1)} \subset \B^{(i)}$ allows us to omit the decoding of certain rows during round $(i-1)$ (and possibly during subsequent rounds).
These ideas were first explored by Bossert \cite{bossert1988decoding}.

Consider the $(i-1)$-th round of decoding ($i \leq k$).
Let $\vec{R}$ and $\vec{R}'$ denote the input to the $i$-th and $(i-1)$-th round of decoding, respectively.
Let $\alpha_j$ denote the reliability weight of the $j$-th row after the decoding with $\B^{i}$.
Let $\widehat{\vec{W}}, \widehat{\vec{V}}$ and $\widehat{\vec{E}}$ denote the guesses for $\vec{W},\ \vec{V}$ and $\vec{E}$ during the $i$-th round, respectively.
Furthermore, let $\vec{e} := \widehat{\vec{V}}^i - \vec{V}^i$, and define
\begin{equation*}
  T := \{ j \in [M]\ |\ \alpha_j = 0 \}
  \qquad
  S := \{ j \in [M]\ |\ e_j \neq 0 \} \setminus T.
\end{equation*}

We have the following observations.
\begin{lemma}
  If $j \notin S \cup T$, then $\vec{R}'_j - \widehat{\vec{E}}_j \in \B^{(i-1)}$.
  \label{lemma:nextround}
\end{lemma}
\begin{IEEEproof}
  Let $\vec{B}^{(i)}$ denote the first $b_i := \sum_{j=1}^i s_j$ rows of $\vec{B}$.
  We know that $\widehat{\vec{W}}_j = \vec{R_j} - \widehat{\vec{E}}_j \in \B^{(i)}$, and $\widehat{\vec{W}} = \widehat{\vec{V}} \vec{B}^{(i)}$.
  Thus,
  \begin{equation*}
    \widehat{\vec{W}} =
    \begin{bmatrix}
      \widehat{\vec{V}}^1 & \dots & \widehat{\vec{V}}^{i-1} & \vec{V}^{i} + \vec{e}
    \end{bmatrix}
    \vec{B}^{(i)}
  \end{equation*}
  and it follows that
  \begin{equation*}
    \vec{R}' =
    \begin{bmatrix}
      \widehat{\vec{V}}^1 & \dots & \widehat{\vec{V}}^{i-1} & \vec{e}
    \end{bmatrix}
    \vec{B}^{(i)} + \widehat{\vec{E}}.
  \end{equation*}
  We have $e_j = 0$ by assumption, and therefore $\vec{R}'_j - \widehat{\vec{E}}_j \in \B^{(i-1)}$.
\end{IEEEproof}
The consequences of Lemma \ref{lemma:nextround} are profound; if $j \notin S \cup T$, then the $j$-th row does not need to be decoded during the next round.
To see this, note that $\vec{R}'_j - \widehat{\vec{E}}_j$ is the unique codeword of $\B^{(i-1)}$ within half the minimum distance of $\vec{R}'_j$.

On the other hand if $j \in S$, then the $j$-th row was incorrectly decoded during the $i$-th round, and hence
\begin{equation*}
  \wt(\vec{E}_j) \geq d_{b,i} - \wt(\widehat{\vec{E}}_j),
\end{equation*}
and hence it is unnecessary to decode the row this round unless
\begin{equation}
  \label{eq:decodecond}
  d_{b,i} - \wt(\widehat{\vec{E}}_j) \leq \lfloor (d_{b,i-1} - 1) /2 \rfloor.
\end{equation}
Finally, if $j \in T$, then the row could not be decoded during the previous round.
Thus, if the error correction capacity of $\B^{(i-1)}$ is not larger than that of $\B^{(i)}$, then we can omit the decoding of this row during this round and instead directly set $\alpha'_j = 0$.
This is the case if $d_{b,i-1} = d_{b,i}$ or, alternatively, if $d_{b,i-1} = d_{b,i} + 1$ and $d_{b,i-1}$ is odd.

These observations suggest the following algorithm for every round except the first, \emph{i.e.} the round with index $k$.
Again consider the $(i-1)$-th round of decoding.
Every row $\vec{R}_j$ such that $j \notin S \cup T$ is not decoded with the inner code $\B^{(i-1)}$, and rows such that $j \in S \cup T$ are only decode if needed.
More precisely, any row $\vec{R}_j$ such that $j \in S$ is decoded only if \eqref{eq:decodecond} is satisfied, while rows with $j \in T$ are decoded only if the error correction capacity of $\B^{(i-1)}$ is larger than that of $\B^{(i)}$.
This modified algorithm is presented as Algorithm \ref{alg:gccimproved}.
\begin{figure}[htbp]
  \removelatexerror
  \begin{algorithm}[H]
    \caption{Improved decoding of generalized concatenated codes}
    \label{alg:gccimproved}
    \begin{algorithmic}[1]
      \Procedure{gccdecode1}{$[\A_{1}, \dots, \A_{k}] \B,\vec{R},\vec{E},S,T$}
	\State $\widehat{\vec{W}}, \widehat{\vec{E}}, \alpha := $ \Call{rowdecode}{$\B, \vec{R},\vec{E},S,T$}
	\State $\widehat{\vec{V}} := \widehat{\vec{W}} \vec{B}^{-1}$
	\State $\vec{V}^k, \vec{e} := $ \Call{gmd-decode}{$\A_k, \widehat{\vec{V}}^k, \alpha$}
	\State $\vec{R}' := \vec{R} - \begin{bmatrix} \vec{0} & \dots & \vec{0} & \vec{V}^k \end{bmatrix} \vec{B}$
	\If{$k > 1$}
	  \State $T' := \{ j \in [M]\ |\ \alpha_j = 0 \}$
	  \State $S' := \{ j \in [M]\ |\ e_j \neq 0 \} \setminus T'$
	  \State $\vec{W}' := $ \Call{gccdecode1}{$[\A_{1}, \dots, \A_{k-1}] \B^{(k-1)}, \vec{R}',\widehat{\vec{E}},S',T'$}
	  \State \Return $\vec{W}' + \begin{bmatrix} \vec{0} & \dots & \vec{0} & \vec{V}^k \end{bmatrix} \vec{B}$
	\Else
	\State \Return $\begin{bmatrix} \vec{0} & \dots & \vec{0} & \vec{V}^k \end{bmatrix} \vec{B}$
	\EndIf
      \EndProcedure
      \State
      \Procedure{rowdecode}{$C, \vec{R}, \vec{E}, S, T$}
	\State Decode every row of $\vec{R}$ with the code $C$.
	\If {$\lfloor (d_{b,k} - 1) / 2 \rfloor > \lfloor (d_{b,k+1} - 1) / 2 \rfloor$}
	  \For {$ j \in T$}
	    \State Decode $\vec{R}_j$ with a decoder for $C$.
	  \EndFor
	\EndIf
	\For {$ j \in T$}
	  \If {$d_{b,k+1} - \wt(\vec{E}_j) \leq \lfloor (d_{b,k} - 1) / 2 \rfloor$}
	    \State Decode $\vec{R}_j$ with a decoder for $C$.
	  \EndIf
	\EndFor
      \EndProcedure
    \end{algorithmic}
  \end{algorithm}
\end{figure}

This modification to the algorithm significantly lowers the number of times the rows have to be decoded.
During the first round we have to decode all the rows.
During the $i$-th round, $i < k$, at most $|S \cup T| < d_{a,i+1}$ rows need to be decoded.
Therefore, the decoder for $\B^{(i)}$, $i < k$, has to be run at most $d_{a,i+1} - 1$ times.
The total number of row decoder invocations is thus at most
\begin{equation*}
  M + \sum_{i=0}^{k-1} \left[ d_{a,k-i} - 1 \right].
\end{equation*}

It is possible to make another small improvement to the algorithm.
Let $\alpha',e',S'$ and $T'$ denote the variables that correspond to $\alpha,e,S$ and $T$ during round $(i-1)$, respectively.
Whenever $j \in S$ and \eqref{eq:decodecond} is not satisfied, then, in Algorithm \ref{alg:gccimproved}, we do not decode the row this round because we already know what it would decode to.
On the other hand, we also know that the row will be decoded to the wrong codeword, and hence -- in order to ease the task of the GMD decoder -- we could directly set $\alpha'_j = 0$.
We should, however, not add $j$ to $T'$.
Instead, we should treat the row as a row without decoding failure, which means that $j \in S'$ if and only if $e'_j \neq 0$.

This improvement does not lower the worst case complexity of the algorithm, but it should lower the number of trials the GMD decoder has to run during the $(i-1)$-th round.

\section{Decoding of matrix-product codes}
\label{sec:decodemp}
Recall that matrix-product codes are GCC, and thus they can be decoded with the same algorithm.
A matrix-product codes is constructed by specifying the generator matrix $\vec{B}$ of the inner code $\B$.
For the rest of this section, suppose $\vec{B}$ is NCS.

Since $\vec{B}$ is NSC, it follows that $\B^{(i)}$ is MDS for all $i$, and hence
$d_{b,i} = N - i + 1$.
We can use this additional structure to lower the upper bound on the number of times the row decoders need to be run.

Since $d_{b,i} = d_{b,i+1} + 1$ for all $i \in [k]$, it follows that the error correction capacity of the row code increases every other round.
Suppose that $d_{b,k}$ is odd.
Then we only need to decode rows during $1 + \lfloor (k-1) / 2 \rfloor$ of $k$ rounds.
Furthermore, the total number of row decoder invocations is at most
\begin{equation*}
  M + \sum_{i=0}^{t-1} \left[ d_{a,k-1-2i} - 1 \right],
\end{equation*}
where $t = \lfloor (k-1) / 2 \rfloor$.
On the other hand, if $d_{b,k}$ is even, then we only need to decode rows during $1 + \lfloor k / 2 \rfloor$ of $k$ rounds.
Furthermore, the total number of row decoder invocations is at most
\begin{equation*}
  M + \sum_{i=0}^{t-1} \left[ d_{a,k-2i} - 1 \right],
\end{equation*}
where $t = \lfloor k / 2 \rfloor$.

\section{Decoding examples}
\label{sec:examples}
We will show how to apply the decoding algorithm in practice by breaking it down for two well known matrix-product codes.
The examples are quite simple but we think they illustrate the algorithm without undue complications.
Throughout this section, let $\sigma_{a,i}$ be an error-and-erasure decoder for the code $\A_i$, and let $\vec{B}_i$ denote the first $i$ rows of $\vec{B}$.

\subsection{Decoding of the $(u\ |\ u + v)$ construction}
Recall that choosing
\begin{equation*}
  \vec{B} =
  \begin{bmatrix}
    1 & 1 \\
    0 & 1
  \end{bmatrix}
\end{equation*}
gives us the $(u\ |\ u + v)$ construction.
Let $d^*$ denote the designed minimum distance of $[\A_1 \A_2] \cdot \vec{B}$.
The naive decoding algorithm for this code is presented in \cite{hernando2013decoding}, and, using the same notation as in Sections \ref{sec:decodecc} and \ref{sec:decodegcc}, is as follows:
\begin{enumerate}
  \item Decode $\vec{R}^2 - \vec{R}^1$ with the decoder for $\A_2$ and denote the result by $\vec{V}^2$.
  \item Decode $\vec{R}^1$ with the decoder for $\A_1$ and denote the result by $\vec{V}^1$.
    If
    \begin{equation*}
      \wt(
      \begin{bmatrix} \vec{V}^1 & \vec{V}^2 \end{bmatrix}
      \cdot \vec{B} - \vec{R}) < d^* / 2,
    \end{equation*}
    then $\vec{V}^1$ is correct so return $(\vec{V}^1, \vec{V}^2)$. Otherwise go to step $3$.
  \item Decode $\vec{R}^2 - \vec{V}^2$ with the decoder for $\A_1$ and denote the result by $\vec{V}^1$.
    Return $(\vec{V}^1, \vec{V}^2)$.
\end{enumerate}
This works for any $\A_1$ and $\A_2$, even if $d_{a,2} \geq 2 d_{a,1}$ does not hold.
The previous requirement is given in \cite{hernando2013decoding}, but it is not necessary.
The worst case complexity of this algorithm is easy to find.
The decoder for $\A_2$ is run once, and the decoder for $\A_1$ is run twice.

Now consider the algorithm for GCC.
$\vec{B}$ is of full rank, and thus $\B$ cannot correct or detect any errors.
Therefore, we can skip the row decoding during round $2$ of decoding.
It follows that $\vec{V}^2$ is recovered exactly as in the naive algorithm.

$\B^{(1)}$ is a repetition code of length $2$, and can thus detect one error.
This means that we only have two reliability classes during round $1$ of decoding.
These correspond to the reliability weights $0$ and $1$ respectively.
Thus the GMD decoder only needs to consider one erasure set, namely $F_1$, when decoding $\vec{R}^1$.
If we choose $\vec{B}_{1}^{-1} = \begin{bmatrix} 1 & 0 \end{bmatrix}^T$, then the algorithm simplifies to
\begin{enumerate}
  \item Let $\widehat{\vec{V}} := \vec{R} \vec{B}^{-1}$, and decode $\widehat{\vec{V}}^2 = \vec{R}^2 - \vec{R}^1$ with the decoder for $\A_2$.
    Denote the result by $\vec{V}^2$, and let $\vec{e}$ be such that $\vec{V}^2 + e = \widehat{\vec{V}}^2$.
  \item Consider $\vec{R}' := \vec{R} - \begin{bmatrix} \vec{0} & \vec{V}^2 \end{bmatrix} \vec{B} $.
    Set $\alpha'_j = 0$ if $e_j \neq 0$, and $\alpha'_j = 1$ otherwise.
  \item Let $\widehat{\vec{V}}' := \vec{R}' \vec{B}_{1}^{-1} = (\vec{R}')^1$, compute
    \begin{equation*}
      \vec{V}^1 := \sigma_{a,1}(\widehat{\vec{V}}', F_1(\alpha'))
      = \sigma_{a,1}( (\vec{R}')^1, F_1(\alpha')),
    \end{equation*}
    and return $(\vec{V}^1, \vec{V}^2)$.
\end{enumerate}
Note that we could as well have chosen $\vec{B}_{1}^{-1} = \begin{bmatrix} 0 & 1 \end{bmatrix}^T$, and then we would instead decode $(\vec{R}')^2 = \vec{R}^2 - \vec{V}^2$ in the last step.
The decoder for $\A_2$ is run once, and the decoder for $\A_1$ is run once.

\subsection{Decoding of the $(u + v + w\ |\ 2u + v\ |\ u)$ construction}
Setting
\begin{equation*}
  \vec{B} =
  \begin{bmatrix}
    1 & 2 & 1 \\
    1 & 1 & 0 \\
    1 & 0 & 0
  \end{bmatrix}
\end{equation*}
gives us the $(u + v + w\ |\ 2u + v\ |\ u)$ construction.
$\B^{(2)}$ has minimum distance $2$ and can thus only detect one error, while $\B^{(1)}$ has minimum distance $3$, which means that it can correct one error.

If we choose
\begin{equation*}
  \vec{B}_{2}^{-1} =
  \begin{bmatrix}
    0 & 1 \\
    0 & 0 \\
    1 & -1
  \end{bmatrix}
  , \quad \text{and} \quad
  \vec{B}_{1}^{-1} =
  \begin{bmatrix}
    0 \\
    0 \\
    1
  \end{bmatrix}
  ,
\end{equation*}
and write the algorithm non-recursively, then the decoding algorithm reduces to:
\begin{enumerate}
  \item Let $\widehat{\vec{V}} := \vec{R} \vec{B}^{-1}$, and decode $\widehat{\vec{V}}^3 = \vec{R}^1 - \vec{R}^2 + \vec{R}^3$ with the decoder for $\A_3$.
    Denote the result by $\vec{V}^3$ and let $\vec{e} := \widehat{\vec{V}}^3 - \vec{V}^3$ denote the corresponding error vector.
  \item Consider $\vec{R}' = \vec{R} - \begin{bmatrix} \vec{0} & \vec{0} & \vec{V}^3 \end{bmatrix} \vec{B}$.
    Compute the reliability weight vector $\alpha'$.
    More precisely, let $\alpha'_i = 0$ if $e_i \neq 0$, and $\alpha'_i = 1$ otherwise.
    Let $\widehat{\vec{V}}' := \vec{R}' \vec{B}_{2}^{-1}$ and decode $(\widehat{\vec{V}}')^2 = (\vec{R}')^1 - (\vec{R}')^3$, \emph{i.e}, set
    \begin{equation*}
      \vec{V}^2 := \sigma_{a,2}((\vec{R}')^1 - (\vec{R}')^3, F_1(\alpha')).
    \end{equation*}
    Let $\vec{e}' := (\widehat{\vec{V}}')^2 - \vec{V}^2$ denote the corresponding error vector.
  \item Consider $\vec{R}'' := \vec{R}' - \begin{bmatrix} \vec{0} & \vec{V}^2 \end{bmatrix} \vec{B}_2$.
    Decode every row such that $\alpha'_i = 0$ or $e'_i \neq 0$ with the decoder for $\B^{(1)}$, and denote the resulting matrix by $\widehat{\vec{W}}''$.
    Let $\alpha''$ denote the corresponding reliability weight vector.
    We set $\alpha''_i = 1$ if the row is not decoded.
    Finally, set $\widehat{\vec{V}}'' := \widehat{\vec{W}}'' \vec{B}_{1}^{-1} = (\widehat{\vec{W}}'')^3$.
  \item Let $\vec{V}^1 := \sigma_{a,1}( \widehat{\vec{V}}'', F_1(\alpha''))$.
    If $\vec{V}^1$ satisfies \eqref{eq:gmdcond}, then return $(\vec{V}^1, \vec{V}^2, \vec{V}^3)$.
    Otherwise continue to the next step.
  \item Let $\vec{V}^1 := \sigma_{a,1}( \widehat{\vec{V}}'', F_2(\alpha''))$ and return $(\vec{V}^1, \vec{V}^2, \vec{V}^3)$.
\end{enumerate}
The decoders for $\A_3$ and $\A_2$ are run once, and the decoder for $\A_1$ is run at most two times.
In addition, the decoder for $\B^{(1)}$ is run at most $d_{a,2} - 1$ times.
Note that the number of reliability classes is not constant during the different rounds of decoding, and as a consequence the definitions for the erasure sets $F_j$ are dependent on the inner code used during that round.

\section{Error-and-erasure decoding of CC and GCC}
\label{sec:erasmp}
The errors-only decoder presented in Section \ref{sec:decodemp} can be turned into an error-and-erasure decoder, by using error-and-erasure decoders for the inner codes, and by using the weighting scheme presented by Wainberg in \cite{wainberg1972error}.
Consider the concatenated code $\A \B$, and let $\vec{R} = \vec{W} + \vec{E}$ denote the received word.
$\widehat{\vec{W}}$ denotes the result after each row has been decoded with the inner code $\B$, and $\widehat{\vec{E}}$ the corresponding error matrix, meaning $\widehat{\vec{W}} + \widehat{\vec{E}} = \vec{R}$.
Let $X_i$ denote the erasure set of the $i$-th row and define
\begin{equation*}
  w_j :=
  \begin{cases}
    2 \wt_{X_j}(\widehat{\vec{E}}_j) + |X_j|,
    & \text{if } 2 \wt_{X_j}(\widehat{\vec{E}}_j) + |X_j| < d_b, \\
    d_b, & \text{in case of decoding failure},
  \end{cases}
\end{equation*}
The reliability weights are given by $\alpha_j := (d_b - w_j) / d_b$.

Wainberg used this weighting scheme to create an error-and-erasure decoder for product codes, but due to the similarities in the expression for the minimum distance of the code, the same weighting scheme works for all concatenated codes.
It is shown in \cite{wainberg1972phd} that if the number of errors $t$ and erasures $s$ is such that $2t + s < d_a d_b$, then -- after the row decoding -- \eqref{eq:gmdcond} is satisfied for every column, and thus the GMD decoder will decode every column correctly.
It is, however, possible to strengthen this statement slightly.

\begin{theorem}
  With this weighting scheme the decoder for the concatenated code can correct all error and erasures patterns that satisfy
  \begin{equation}
    \label{eq:erasurecond}
    \sum_{i \in [M]} \min\left\{ 2 \wt_{X_j}(\vec{E}_i) + |X_i|,\ 2 d_b \right\} < d_a d_b.
  \end{equation}
  \label{thm:erasurecorrect}
\end{theorem}
\begin{IEEEproof}
  Let $I_C \subset [M]$ denote all rows that were correctly decoded, and define $I_E := [M] \setminus I_C$.
  Suppose $j \in I_C$.
  Then
  \begin{IEEEeqnarray*}{rCl}
    \min \left\{ 2 \wt_{X_j}(\vec{E}_j) + |X_j|,\ 2 d_b \right\}
    &=& 2 \wt_{X_j}(\vec{E}_j) + |X_j|
    = 2 \wt_{X_j}(\widehat{\vec{E}}_j) + |X_j| \\
    &=& d_b - \left( d_b - w_j \right)
    = (1 - \alpha_j) d_b.
  \end{IEEEeqnarray*}
  If $j \in I_E$, and we don't have a decoding failure for that row, then
  \begin{IEEEeqnarray*}{rCl}
    \min \left\{ 2 \wt_{X_j}(\vec{E}_j) + |X_j|,\ 2 d_b \right\}
    & \geq & 2 d_b - \left(2 \wt_{X_j}(\widehat{\vec{E}}_j) + |X_j| \right) \\
    &=& d_b + \left( d_b - w_j \right)
    = (1 + \alpha_j) d_b.
  \end{IEEEeqnarray*}
  On the other and, if we have a decoding failure for the $j$-th row, then $\alpha_j = 0$, and hence
  \begin{IEEEeqnarray*}{rCl}
    \min \left\{ 2 \wt_{X_j}(\vec{E}_j) + |X_j|,\ 2 d_b \right\}
    & \geq & d_b = (1 + \alpha_j) d_b.
  \end{IEEEeqnarray*}

  Now, assume that \eqref{eq:erasurecond} holds.
  Then,
  \begin{IEEEeqnarray*}{rCl}
    \sum_{i \in I_C} (1 - \alpha_i) d_b
    + \sum_{i \in I_E} (1 + \alpha_i) d_b
    & \leq &
    \sum_{j \in [M]} \min \left\{ 2 \wt_{X_j}(\vec{E}_j) + |X_j|,\ 2 d_b \right\}
    < d_a d_b,
  \end{IEEEeqnarray*}
  which shows that \eqref{eq:gmdcond} holds for all columns.
  Hence a GMD decoder will decode every column correctly.
\end{IEEEproof}

We see that
\begin{equation*}
  \sum_{i \in [M]} \min\left\{ 2 \wt_{X_j}(\vec{E}_i) + |X_i|,\ 2 d_b \right\}
  \leq \sum_{i \in [M]} 2 \wt_{X_j}(\vec{E}_i) + |X_i|
\end{equation*}
and hence we have the following corollary.
\begin{corollary}
  With this weighting scheme the decoder for the concatenated code can correct all error and erasures patterns that satisfy
  $2 t + s < d_a d_b$, where $t$ is the number of errors and $s$ the numbers of erasures.
\end{corollary}

We call a row \emph{bursty} if $2 \wt_{X_j}(\vec{E}_i) + |X_i| \geq d_b$.
We will now prove the statement analogous to Corollary \ref{cor:bursty}.
\begin{corollary}
  Let $b$ denote the number of bursty rows, and $t$ and $s$ the remaining number of errors and erasures, respectively.
  The concatenated code decoder can decode any error pattern such that
  \begin{equation*}
    2 t + s + 2 b d_b < d_a d_b.
  \end{equation*}
\end{corollary}
\begin{IEEEproof}
  If the $i$-th row is not bursty, then
  \begin{equation*}
    \min \left\{ 2 \wt_{X_j}(\vec{E}_i) + |X_i|,\ 2 d_b \right\}
    = 2 \wt_{X_j}(\vec{E}_i) + |X_i|.
  \end{equation*}
  If the $i$-th row is bursty, then
  \begin{equation*}
    \min \left\{ 2 \wt_{X_j}(\vec{E}_i) + |X_i|,\ 2 d_b \right\}
    \leq 2 d_b,
  \end{equation*}
  and hence the result follows.
\end{IEEEproof}

The enhancement for errors-only decoding to error-and-erasure decoding does not come for free; we have increased the number of reliability classes to $d_b + 1$, and thus the GMD decoder needs to run at most
\begin{equation*}
  m := \min \left\{ d_b, \left\lfloor \frac{d_a + 1}{2} \right\rfloor \right\}
\end{equation*}
trials for each $\widehat{\vec{V}}^i,\ i \in [k]$.

By using the same technique as in Section \ref{sec:decodecc} one can show that
all the $\widehat{\vec{V}}^i$ can be decoded in at most $k + m - 1$ trials.
Thus, the algorithm runs the decoder for the inner code $M$ times, and the decoder for the outer code at most $k + m - 1$ times.

Note that the weighting scheme for error-and-erasure decoding reduces to the one presented in Section \ref{sec:decodecc} if the parameter $X_j$ is neglected.
Thus the error-and-erasure algorithm reduces to the errors-only algorithm if there are no erasures.
Finally, the same weighting scheme can be used for error-and-erasure decoding of GCC.

\section{Decoding beyond the designed minimum distance}
\label{sec:beyondmd}
There are two obvious ways to increase the number of error patterns that the decoding algorithms can correct; allowing decoding beyond the GMD when decoding $\widehat{\vec{V}}^i$, and not settling for minimum distance decoding of the inner code(s).
We will consider each of these techniques in turn.

Decoding beyond the GMD, that is, choosing the trial that minimizes the left hand side of \eqref{eq:gmdcond}, was considered in the context of product codes in \cite{blomqvist2020}.
The decoding performance over a $q$-ary symmetric channel improved by several orders of magnitude.
This is hardly surprising, as similar improvement in performance is achieved when allowing decoding beyond the minimum distance of a code.

Decoding beyond the GMD should yield similar improvement with concatenated codes, and probably also with GCC.
It is easy to see that a decoder that decodes beyond the GMD corrects all error patterns that a decoder that decodes up to the GMD can correct.

Using a decoder for the inner code(s) that decodes beyond the minimum distance was first considered by Bossert in the context of GCC \cite{bossert1988decoding}.
For simplicity of presentation we will restrict the disposition to concatenated codes, as the extension to the case of GCC is straight forward.

Consider the concatenated code $\A \B$, and let the reliability weights be defined as in Section \ref{sec:decodecc}.
The GMD decoder for the outer code will first consider all rows where decoding with the inner code $\B$ failed as erased.
During the next trial the rows that appear to have error weight $t := \lfloor (d_b - 1) / 2 \rfloor$ will also be considered as erasures.
The rows that appear to have error weight $t - 1$ are marked as erasures during the following trial, and so forth.

If the decoder for the inner code can correct a row with more than $t$ errors, say $t'$, then we can instead erase the rows such that $\wt(\widehat{\vec{E}}_j) = t'$ during the second trial.
Then we erase rows such that $\wt(\widehat{\vec{E}}_j) = t' - 1$ during the third trial, and so forth.
We do however, use the reliability weights as defined in Section \ref{sec:decodecc} when checking if the current trial decoded correctly.
In other words, we define
\begin{equation*}
  w_j :=
  \begin{cases}
    2 \wt(\widehat{\vec{E}}_j), & \text{if }  2 \wt(\widehat{\vec{E}}_j) < d_b, \\
    d_b, & \text{otherwise}.
  \end{cases}
\end{equation*}
Another way to view this modification is as a GMD decoder where, instead of the erasure set $F_1$, we have a chain of $t' - t + 1$ subsets of $F_1$, and we try decoding with these erasure sets before continuing with $F_2$.

It is clear that this modified decoder for the outer code $\A$ can correct all error patterns that are correctable by the normal GMD decoder; the modified decoder will try decoding with a collection of erasure sets $\Gamma_1 \supset \Gamma_2$, where $\Gamma_2$ is the erasure set collection that the GMD decoder will use.

\bibliographystyle{IEEEtran}
\bibliography{mpdecode}

\end{document}